\documentclass[journal=jctcce,manuscript=article,layout=twocolumn]{achemso}
\setkeys{acs}{articletitle = true, chaptertitle = true, doi = true}

\usepackage[T1]{fontenc}
\usepackage{mathptmx}

\makeatletter
\let\l@addto@macro\relax
\makeatother
\usepackage[fontsize=9pt]{scrextend}

\usepackage{bm}
\usepackage{amssymb}
\usepackage{amsmath}
\usepackage{hyperref}
\hypersetup{colorlinks=true, linkcolor=black, urlcolor=black, citecolor=black,
plainpages=false, bookmarksnumbered=true}
\usepackage{xurl}
\usepackage[utf8]{inputenc}
\usepackage{booktabs}

\newcommand*{\Tr}{\operatorname{Tr}}
\newcommand*{\mat}[1]{\mathbf{#1}}
\newcommand*{\Abse}{\bm{\mathcal{A}}}
\newcommand*{\Bbse}{\bm{\mathcal{B}}}
\newcommand*{\Xp}{\mat{X}^{+}}
\newcommand*{\Xm}{\mat{X}^{-}}
\newcommand*{\eqp}{\varepsilon^{\mathrm{QP}}}

\author{Christof Holzer}
\affiliation{Institute of Quantum Materials and Technologies, Karlsruhe
Institute of Technology (KIT), Kaiserstra\ss{}e 12, 76131 Karlsruhe,
Germany}
\email{holzer@kit.edu}
\title{Analytic $GW$-Bethe--Salpeter Gradients with Quasiparticle Response in the Resolution of the Identity}

\begin{document}

\begin{abstract}
Analytic nuclear gradients for excited states from the
Bethe--Salpeter equation (BSE) in the static screening approximation
are presented within this work.
The response of the screened interaction to nuclear displacements is
treated within the resolution-of-the-identity (RI) approximation, in close
analogy to RI random phase approximation gradients, and is folded into the
Lagrangian formalism of time-dependent density functional theory gradients.
Quasiparticle gradients enter via the full analytic derivative of the corrections 
for RI contour-deformation $G_0W_0$ and ev$GW$, also taking into account that 
incomplete quasiparticle spectra are typically supplied. 
Comparisons to earlier approaches using a geometry-independent 
fixed quasiparticle correction are made. For the latter,
the orbital-rotation terms are derived exactly, too. 
The implemented gradient scales as $\mathcal{O}(N^4)$,
and the three-index integrals are processed in batches with a memory demand of
$\mathcal{O}(N_\mathrm{aux}^2 + N_\mathrm{occ}^2 N_\mathrm{aux})$.
Excited-state geometries of a set of small molecules are compared to 
EOM-CCSD references and to TD-DFT. It is shown that the full quasiparticle 
gradient removes most of the systematic underestimation of the bond lengths
in singlet excited states. A comparison along the C--C stretch of acetylene 
shows that the neglected derivative of the quasiparticle corrections is modest, 
but becomes erratic for $G_0W_0$ near poles of the self-energy, which ev$GW$ avoids.
\end{abstract}

%%%%%%%%%%%%%%%%%%%%%%%%%%%%%%%%%%%%%%%%%%%%%%%%%%%%%%%%%%%%%%%%%%%%%
\section{Introduction}
%%%%%%%%%%%%%%%%%%%%%%%%%%%%%%%%%%%%%%%%%%%%%%%%%%%%%%%%%%%%%%%%%%%%%
The $GW$ approximation\cite{Hedin1965} combined with the Bethe--Salpeter
equation (BSE)\cite{Salpeter1951,Strinati1988,Rohlfing2000,Onida2002} has become an established
tool for molecular excitation energies, in particular for charge-transfer
and Rydberg states, with a BSE step at a cost comparable to time-dependent
density functional theory (TDDFT).\cite{Blase2018,Blase2020}
Charge-transfer excitations, which are underestimated by TDDFT with standard functionals, are described
well by the BSE.\cite{Blase2011CT,Duchemin2012}
It is now a viable method to describe
excited states\cite{Bruneval2015,Jacquemin2015,Jacquemin2017,Gui.Holzer.ea:Accuracy.2018}
as well as molecular properties,\cite{Holzer2025Guide,Holzer2017SAPT,Franzke2022NMR,
Holzer.Klopper:Ionized.2019,Holzer2018cBSE}
including excited-state properties and non-linear response\cite{Rauwolf2024,Himmelsbach2024}
and real-time propagation approaches.\cite{Attaccalite2011,Marek2025} 
To further expand the reach of BSE towards excited-state structures, 
emission energies and photochemical paths, however, requires
nuclear gradients. Early work on excited-state forces in the $GW$-BSE framework approximated
derivatives of the quasiparticle (QP) Hamiltonian by those of the density
functional theory (DFT) Hamiltonian,\cite{IsmailBeigi2003,DelGrande2025}
molecular excited-state geometries were optimized with numerical
gradients,\cite{Ma2010,Caylak2021} and Lagrangian Z-vector formulations were
assessed for molecular properties.\cite{VillalobosCastro2023}
Fully analytic gradients of BSE@$G_0W_0$ starting from Hartree--Fock (HF)
have been reported recently,\cite{Toelle2025BSE} building on analytic
$G_0W_0$ gradients,\cite{Toelle2025GW} with a computational cost that scales
as $\mathcal{O}(N^6)$ to $\mathcal{O}(N^7)$.
A plane-wave implementation of analytic BSE forces for solids with hundreds of atoms has also appeared,
which approximates the derivative of the QP Hamiltonian by that of the Kohn--Sham Hamiltonian
and neglects the derivative of the screened interaction with respect to the orbitals.\cite{Jin2026}

Here, I present analytic BSE gradients in the RI formulation of
\textsc{Turbomole},\cite{Balasubramani.Chen.ea:TURBOMOLE.2020,Franzke.Holzer.ea:TURBOMOLE.2023} which is used for
BSE excitation energies.\cite{Krause2017,Holzer2018cBSE}
The derivative of the statically screened interaction with respect to the
orbitals, orbital energies and integrals is formulated with the techniques
of RI random phase approximation (RPA) energies and gradients\cite{Eshuis2010,Burow2014} and
embedded in the Lagrangian of TDDFT gradients.\cite{Furche2002,Rappoport2005}
QP energies enter with geometry-independent $GW$ corrections.
The additional terms arising from the geometry dependence of these
corrections are derived in Section~\ref{sec:gw}.

%%%%%%%%%%%%%%%%%%%%%%%%%%%%%%%%%%%%%%%%%%%%%%%%%%%%%%%%%%%%%%%%%%%%%
\section{Theory}
\label{sec:theory}
%%%%%%%%%%%%%%%%%%%%%%%%%%%%%%%%%%%%%%%%%%%%%%%%%%%%%%%%%%%%%%%%%%%%%
The formulation assumes closed-shell references with real canonical orbitals.
Indices $i,j,k$ denote occupied, $a,b,c$ virtual and $p,q,r,s$ general
molecular orbitals (MOs), and $P,Q,R,S$ auxiliary basis functions.
A generalization from spatial orbitals to spin
orbitals is straightforward.
Three-index integrals in the orthonormalized auxiliary basis are
\begin{equation}
  \begin{split}
    B^P_{pq} &= \sum_R (pq|R)\,[\mat{V}^{-1/2}]_{RP},\\
    V_{RS} &= (R|S).
  \end{split}
\end{equation}

\subsection{Bethe--Salpeter equation}
The BSE excitation energies $\Omega$ follow from
\begin{equation}
(\Abse-\Bbse)(\Abse+\Bbse)|\mat{X}+\mat{Y}\rangle
= \Omega^2|\mat{X}+\mat{Y}\rangle
\end{equation}
with, for singlet excitations,
\begin{align}
  \mathcal{A}_{ia,jb} &= \delta_{ij}\delta_{ab}(\varepsilon_a-\varepsilon_i)
             + 2(ia|jb) - W_{ij,ab},\\
  \mathcal{B}_{ia,jb} &= 2(ia|jb) - W_{ib,ja}.
\end{align}
For triplets, the Coulomb term $2(ia|jb)$ is absent.
The orbital energies $\varepsilon_p$ in $\Abse$ are either Kohn--Sham (KS)
or QP energies.
Within RI, the statically screened interaction reads
\begin{equation}
  \begin{split}
    W_{pq,rs} &= \sum_{PQ} B^P_{pq} M_{PQ} B^Q_{rs},\\
    \mat{M} &= (\mat{1}-\bm{\Pi})^{-1},
  \end{split}
\end{equation}
with the static response matrix
\begin{equation}
  \begin{split}
    \Pi_{PQ} &= -4\sum_{kc}\frac{B^P_{kc}B^Q_{kc}}{\Delta_{kc}},\\
    \Delta_{kc} &= \varepsilon_c-\varepsilon_k,
  \end{split}
  \label{eq:pi}
\end{equation}
where the orbital energies in $\bm{\Pi}$ may differ from those in
$\Abse$.
In the Tamm--Dancoff approximation (TDA), $\Bbse$ and $\mat{Y}$ are
set to zero.

\subsection{Lagrangian}
Following Furche and Ahlrichs,\cite{Furche2002} the excitation energy is
written as
\begin{equation}
  \Omega = \tfrac12\left[\langle\Xp|\Abse+\Bbse|\Xp\rangle
         + \langle\Xm|\Abse-\Bbse|\Xm\rangle\right],
\end{equation}
with $\mat{X}^{\pm}=\mat{X}\pm\mat{Y}$ and
$\langle\Xp|\Xm\rangle = 1$.
Stationarity with respect to the MO coefficients is enforced by the
Lagrange multipliers $\mat{Z}$ and $\bm{\Theta}$,\cite{Handy1984} such that the
gradient follows from the relaxed difference density
$\mat{P} = \mat{T}+\mat{Z}$, the energy-weighted density $\bm{\Theta}$ and
two-particle terms.\cite{Furche2002,Rappoport2005}
Here $\mat{T}$ is the unrelaxed excitation difference density, before
adding the screening contribution, with blocks
\begin{align}
  T_{ij} &= -\tfrac12\sum_a
    \left(X^+_{ia}X^+_{ja}+X^-_{ia}X^-_{ja}\right),\\
  T_{ab} &= \tfrac12\sum_i
    \left(X^+_{ia}X^+_{ib}+X^-_{ia}X^-_{ib}\right),\\
  T_{ia} &= T_{ai}=0.
  \label{eq:difference-density}
\end{align}
All terms of $\Omega$ that also occur in TDDFT, i.e., the orbital energy
differences and the Coulomb term, are treated as
in RI-TDDFT gradients.\cite{Rappoport2005,Furche2002}
The Z-vector equation contains the full exchange--correlation kernel of the
ground state.
The following derivation therefore focuses on the screened exchange term.

\subsection{Screened exchange term}
Collecting the terms of $\Omega$ that contain the screened interaction $W$ gives
\begin{equation}
  \Omega_W = \sum_{PQ} M_{PQ}\,\Gamma^W_{PQ},
\end{equation}
\begin{equation}
  \Gamma^W_{PQ} = -\tfrac12\left[
    \Tr(\mat{B}^P_{\mathrm{oo}}\mat{R}^Q)
  + \Tr(\mat{K}^P\mat{K}^Q)
  - \Tr(\mat{L}^P\mat{L}^Q)\right],
  \label{eq:gamw}
\end{equation}
symmetrized in $P,Q$, with the occupied--occupied matrices
\begin{equation}
  \begin{split}
    \mat{R}^Q &= \Xp\mat{B}^Q_{\mathrm{vv}}\mat{X}^{+\mathsf{T}}
              + \Xm\mat{B}^Q_{\mathrm{vv}}\mat{X}^{-\mathsf{T}},\\
    \mat{K}^P &= \mat{B}^P_{\mathrm{ov}}\mat{X}^{+\mathsf{T}},\\
    \mat{L}^P &= \mat{B}^P_{\mathrm{ov}}\mat{X}^{-\mathsf{T}},
  \end{split}
\end{equation}
where $\mat{X}^\pm$ are arranged as $N_\mathrm{occ}\times N_\mathrm{vir}$
matrices.
With $\mathrm{d}\mat{M} = \mat{M}\,\mathrm{d}\bm{\Pi}\,\mat{M}$, the
derivative separates into a frozen-$\mat{M}$ part and the response of the
screening,
\begin{equation}
  \begin{split}
    \mathrm{d}\Omega_W &= \sum_{PQ} M_{PQ}\,\mathrm{d}\Gamma^W_{PQ}
    + \Tr\big(\bar{\bm{\Gamma}}\,\mathrm{d}\bm{\Pi}\big),\\
    \bar{\bm{\Gamma}} &= \mat{M}\bm{\Gamma}^W\mat{M}.
  \end{split}
\end{equation}
The frozen-$\mat{M}$ part has the structure of the exact-exchange term of
RI hybrid TDDFT gradients with one set of integrals replaced by the screened
integrals $\tilde{\mat{B}} = \mat{M}\mat{B}$.
The screening response is obtained analogously to the frequency-integrated
terms of RI-RPA gradients.\cite{Burow2014}
Differentiating equation~\eqref{eq:pi} and defining
\begin{equation}
  C^P_{kc} = \frac{1}{\Delta_{kc}}\sum_Q \bar{\Gamma}_{PQ} B^Q_{kc}
\end{equation}
yields
\begin{equation}
  \Tr\big(\bar{\bm{\Gamma}}\,\mathrm{d}\bm{\Pi}\big)
  = -8\sum_{Pkc} C^P_{kc}\,\mathrm{d}B^P_{kc}
  + \sum_{pq} D^{\Pi}_{pq}\,\mathrm{d}F_{pq},
\end{equation}
where the orbital energy derivatives were generalized to the derivative of
the Fock matrix $F_{pq}$ with the screening response densities
\begin{align}
    D^{\Pi}_{ab} &= 4\sum_{kP} \frac{C^P_{ka}B^P_{kb}}{\Delta_{kb}},\\
    D^{\Pi}_{ij} &= -4\sum_{cP} \frac{C^P_{ic}B^P_{jc}}{\Delta_{jc}},
\end{align}
both symmetrized.
$\mat{D}^{\Pi}$ is added to the unrelaxed difference density $\mat{T}$ and
thus enters the Z-vector equation, the energy-weighted density and the
one-electron part of the gradient in the same way as $\mat{T}$.

All derivatives of the three-index integrals are collected in
$g^P_{pq} = \partial\Omega_W/\partial B^P_{pq}$, comprising the
frozen-$\mat{M}$ contributions of equation~\eqref{eq:gamw} and
$g^P_{kc} \leftarrow -8C^P_{kc}$.
The orbital response of $B^P_{pq}$ gives the matrix $\mat{Q}$,
\begin{equation}
  Q_{rp} = \sum_{Pq} B^P_{rq}\left(g^P_{pq}+g^P_{qp}\right),
\end{equation}
which enters the right-hand side of the Z-vector equation as
$Q_{ai}-Q_{ia}$ and the energy-weighted density.\cite{Furche2002}
The integral derivatives at fixed MOs follow from
\begin{align}
  \Gamma^R_{\mu\nu} &= \sum_P [\mat{V}^{-1/2}]_{RP}
     \sum_{pq} C_{\mu p}\, g^P_{pq}\, C_{\nu q},\\
  \Gamma_{RS} &= -\tfrac12 \sum_{PQ} [\mat{V}^{-1/2}]_{RP}\,
     G_{PQ}\,[\mat{V}^{-1/2}]_{SQ},
\end{align}
with $G_{PQ} = \sum_{pq} g^P_{pq}B^Q_{pq}$ symmetrized, which are contracted
with $(\mu\nu|R)^{\xi}$ and $(R|S)^{\xi}$, respectively.\cite{Haettig2003,Burow2014}
The simple form of $\Gamma_{RS}$ holds because $\Omega$ is invariant under
orthogonal transformations of the auxiliary index, as $\mat{M}$ is
defined in the same orthonormalized basis.

\section{Geometry dependence of the $GW$ corrections}
\label{sec:gw}
%%%%%%%%%%%%%%%%%%%%%%%%%%%%%%%%%%%%%%%%%%%%%%%%%%%%%%%%%%%%%%%%%%%%%
The gradient of the BSE energy for given QP energies follows from the
RI-K and RI-RPA gradient techniques of Section~\ref{sec:theory}.
The derivatives of the $GW$ corrections of the RI-based $G_0W_0$ and ev$GW$
variants require no new integral types: the screening response has the structure
of the frequency-integrated terms of RI-RPA gradients, supplemented by
frequency-dependent weights, contractions over all orbital pairs and, for the
residue terms, screening at real frequencies.
Special care is required because many $GW$ calculations do not correct all orbitals, 
but only a small subset.
Let $\mathrm{o}$ denote the lowest and $\mathrm{v}$ the highest orbital
for which a correction is computed, with $i_{\mathrm{HOMO}}$ and
$a_{\mathrm{LUMO}}$ the frontier orbitals.
The corrected set is then $\mathcal{I}=\mathcal{I}_{\mathrm{occ}}\cup\mathcal{I}_{\mathrm{vir}}$
with $\mathcal{I}_{\mathrm{occ}}=\{\mathrm{o},\dots,i_{\mathrm{HOMO}}\}$
and $\mathcal{I}_{\mathrm{vir}}=\{a_{\mathrm{LUMO}},\dots,\mathrm{v}\}$.
All other orbitals take the correction of the closest corrected one,
$c_i=c_{\mathrm{o}}$ for the occupied orbitals $i<\mathrm{o}$ and
$c_a=c_{\mathrm{v}}$ for the virtual orbitals $a>\mathrm{v}$.
Sums over $p$ below run over all orbitals if every $c_p$ is computed.
Otherwise they run over $\mathcal{I}$, and the weights of the shifted
orbitals are added to those of the orbitals they follow,
\begin{equation}
  \begin{split}
    \tilde w_{\mathrm{o}} &= w_{\mathrm{o}}+\sum_{i<\mathrm{o}}w_i,\\
    \tilde w_{\mathrm{v}} &= w_{\mathrm{v}}+\sum_{a>\mathrm{v}}w_a,
  \end{split}
  \label{eq:folded-weights}
\end{equation}
while $\tilde w_p=w_p$ for all other $p\in\mathcal{I}$.
With $\eqp_p = \varepsilon_p + c_p$, the total derivative is
\begin{equation}
  \begin{split}
    \frac{\mathrm{d}\Omega}{\mathrm{d}\xi}
    &= \left.\frac{\mathrm{d}\Omega}{\mathrm{d}\xi}\right|_{\mat{c}}
    + \sum_p w_p\,\frac{\mathrm{d}c_p}{\mathrm{d}\xi},\\
    w_p &= \frac{\partial\Omega}{\partial\eqp_p},
  \end{split}
\end{equation}
where the first term is given in Section~\ref{sec:qp}.
$\Omega$ is not stationary with respect to the QP energies, 
so the second term does not vanish in general.
The weights are the diagonals of the densities already available,
$w_p = T_{pp} + D^{\Pi}_{pp}$, where $\mat{T}$ here denotes the BSE
difference density before adding $\mat{D}^{\Pi}$.
The $T_{pp}$ term contributes if QP energies enter $\Abse$ and
$D^{\Pi}_{pp}$ if they enter $\bm{\Pi}$.
In the TDA, $\sum_a T_{aa} = -\sum_i T_{ii} = 1$, while
$\sum_a D^{\Pi}_{aa} = -\sum_i D^{\Pi}_{ii}$.
Consequently, if QP energies enter both $\Abse$ and $\bm{\Pi}$ and
$c_i=c_{\mathrm{o}}$, $c_a=c_{\mathrm{v}}$, then
\begin{equation}
  \sum_p w_p\,\mathrm{d}c_p
  = \left(1+\sum_a D^{\Pi}_{aa}\right)
    \mathrm{d}(c_{\mathrm{v}}-c_{\mathrm{o}}).
  \label{eq:qp-gap-weight}
\end{equation}
The coefficient reduces to unity if screening does not depend on the QP
corrections or its contribution vanishes.
Without the TDA, $\sum_a T_{aa} = -\sum_i T_{ii}
= \sum_{ia}(X_{ia}^2+Y_{ia}^2) = 1+2\sum_{ia}Y_{ia}^2$, and the unit
coefficient in equation~\eqref{eq:qp-gap-weight} is replaced by this sum.
Screening can give non-zero weights to orbitals with no amplitude in the
BSE excitation.
With the weights frozen at a reference geometry $\xi_0$, define
\begin{equation}
  \Lambda(\xi;\xi_0) = \sum_p w_p(\xi_0)\,c_p(\xi).
\end{equation}
Its derivative at $\xi_0$ supplies the missing contribution without
derivatives of $\mat{w}$.
A single additional Lagrangian for this weighted sum contributes to the
same orbital Z-vector equation, independent of the number of QP energies.
The fixed-correction approximation of Section~\ref{sec:qp} neglects this
contribution.
The following derivation assumes the contour-deformation (CD) variants.
For analytic continuation,\cite{Duchemin.Blase:Robust.2020,Panades-Barrueta.Golze:Accelerating.2023,
Kehry.Klopper.ea:Robust.2023}
the fitted continuation function must also be differentiated.
For a corrected window $\mathcal{I}$, the weights $\tilde w_p$ of
equation~\eqref{eq:folded-weights} replace $w_p$, and $\sigma_p$ is evaluated for
$p\in\mathcal{I}$ only.
In ev$GW$, the shifted energies also enter $G$ and $W$, and their dependence
on $c_{\mathrm{o}}$ and $c_{\mathrm{v}}$ is part of the Jacobian
in equation~\eqref{eq:evgw-jacobian}.

\subsection{Gradient correction for RI-CD-$G_0W_0$}
\label{sec:qp_g0w0}
Define the real self-energy correction
\begin{equation}
  \sigma_p(\omega;\xi) = \Sigma^{\mathrm{x}}_{pp}(\xi)
  + \operatorname{Re}\Sigma^{\mathrm{c}}_{pp}(\omega;\xi)
  - v^{\mathrm{xc}}_{pp}(\xi).
\end{equation}
In $G_0W_0$, $G_0$ and $W_0$ are constructed with KS orbitals and energies,
and the diagonal QP equation is
\begin{equation}
  \eqp_p = \varepsilon_p + \sigma_p(\eqp_p;\xi).
\end{equation}
If a locally smooth QP function is assumed, differentiation gives
\begin{equation}
  \begin{split}
    \sum_p w_p\frac{\mathrm{d}c_p}{\mathrm{d}\xi}
    ={}& \sum_p v_p\left.\partial_\xi\sigma_p(\omega;\xi)
         \right|_{\omega=\eqp_p}\\
       &+ \sum_p (v_p-w_p)\frac{\mathrm{d}\varepsilon_p}{\mathrm{d}\xi},
  \end{split}
  \label{eq:g0w0}
\end{equation}
with $v_p = Z_p w_p$ and
$Z_p = [1-\partial_\omega\sigma_p(\omega;\xi)]^{-1}_{\omega=\eqp_p}$.
Here $\partial_\xi$ holds only the external frequency $\omega$ fixed:
it includes the response of the KS orbitals, KS energies and integrals
inside $G_0$ and $W_0$.
No derivative of $Z_p$ is required in equation~\eqref{eq:g0w0}.

If the QP equation is solved in its linearized form, which
is done in this work, the coefficients $c_p$ are obtained as
\begin{equation}
  \begin{split}
    c_p &= Z_p^{\mathrm{lin}}\sigma_p(\varepsilon_p;\xi),\\
    Z_p^{\mathrm{lin}} &=
    \left[1-\partial_\omega\sigma_p\right]^{-1}_{\omega=\varepsilon_p}.
  \end{split}
\end{equation}
The derivative includes the response of the renormalization factor and
of its expansion point:
\begin{equation}
  \begin{split}
    \frac{\mathrm{d}c_p}{\mathrm{d}\xi}
    ={}& Z_p^{\mathrm{lin}}\partial_\xi\sigma_p
      +(Z_p^{\mathrm{lin}}-1)\frac{\mathrm{d}\varepsilon_p}{\mathrm{d}\xi}\\
      &+c_pZ_p^{\mathrm{lin}}\left[
        \partial_\xi\partial_\omega\sigma_p
        +\partial_\omega^2\sigma_p
         \frac{\mathrm{d}\varepsilon_p}{\mathrm{d}\xi}\right],
  \end{split}
  \label{eq:qp-linearized-gradient}
\end{equation}
with all derivatives of $\sigma_p$ evaluated at $\omega=\varepsilon_p$.
With $v_p=Z_p^{\mathrm{lin}}w_p$ and $u_p=c_pv_p$, the weighted sum becomes
\begin{equation}
  \begin{split}
    \sum_p w_p\frac{\mathrm{d}c_p}{\mathrm{d}\xi}
    ={}& \sum_p v_p\,\partial_\xi\sigma_p
       + \sum_p u_p\,\partial_\xi\partial_\omega\sigma_p\\
       &+ \sum_p\left(v_p-w_p+u_p\,\partial_\omega^2\sigma_p\right)
         \frac{\mathrm{d}\varepsilon_p}{\mathrm{d}\xi}.
  \end{split}
  \label{eq:qp-linearized-weighted}
\end{equation}
The first term and the $(v_p-w_p)$ part of the last term have the form of equation~\eqref{eq:g0w0},
with $Z_p^{\mathrm{lin}}$ evaluated at the KS energy.
The remaining terms differentiate the frequency dependence of the screening response.
In the imaginary-axis term, the kernel $x/(x^2+\nu^2)$ is replaced by its first
derivative with respect to $\omega$ for $u_p\,\partial_\xi\partial_\omega\sigma_p$
and by its second derivative for $\partial_\omega^2\sigma_p$.
In the residue term, $\partial_zW^{\mathrm{c}}_{ps,ps}$ and its fixed-frequency derivative appear
in the first case, and the second derivative $\partial_z^2W^{\mathrm{c}}_{ps,ps}$ in the second.
These terms need two further sets of screening response densities with differentiated frequency
kernels but no new integral types.
The expressions below give the terms common to equations~\eqref{eq:g0w0} and
\eqref{eq:qp-linearized-weighted}, i.e., the first term and the $(v_p-w_p)$ part,
with $v_p=Z_p^{\mathrm{lin}}w_p$ and the external frequency $\omega=\varepsilon_p$ in the linearized case.
The additional terms of equation~\eqref{eq:qp-linearized-weighted} are built from the same ingredients and
are included in the implementation.

The exchange part, $\Sigma^{\mathrm{x}}_{pp} = -\sum_{iP}(B^P_{pi})^2$, and
$v^{\mathrm{xc}}_{pp}$ are contracted with the weighted density
$\sum_p v_p|\varphi_p\rangle\langle\varphi_p|$. The latter requires the
exchange--correlation kernel for the orbital response and the explicit
derivative, as in TDDFT gradients.\cite{Furche2002}
In the RI formulation, define the auxiliary vector
$\mat{b}_{ps}=(B^P_{ps})_P$ and
\begin{align}
  W^{\mathrm{c}}_{ps,ps}(z)
    &= \mat{b}_{ps}^{\mathsf{T}}[\mat{M}(z)-\mat{1}]\mat{b}_{ps},\\
  \mat{M}(z) &= [\mat{1}-\bm{\Pi}(z)]^{-1},\\
  \Pi_{PQ}(z) &= -4\sum_{kc} B^P_{kc}B^Q_{kc}
    \frac{\Delta_{kc}}{\Delta_{kc}^2-z^2}.
  \label{eq:gw-pi-complex}
\end{align}
The real part of the CD correlation self-energy is obtained from an
imaginary-axis term and residues,\cite{Golze2018,Golze2019}
\begin{equation}
  \operatorname{Re}\Sigma^{\mathrm{c}}_{pp}(\omega)
  = \Sigma^{\mathrm{c,I}}_{pp}(\omega)
    +\operatorname{Re}\Sigma^{\mathrm{c,R}}_{pp}(\omega).
\end{equation}
For the real part needed in the QP equation, the imaginary-axis term is
\begin{equation}
  \begin{split}
    \Sigma^{\mathrm{c,I}}_{pp}(\omega)
    ={}& -\frac{1}{\pi}\int_0^\infty\!\mathrm{d}\nu\,
    \sum_s W^{\mathrm{c}}_{ps,ps}(\mathrm{i}\nu)\\
    &\hspace{12mm}\times
    \frac{\omega-\varepsilon_s}{(\omega-\varepsilon_s)^2+\nu^2}.
  \end{split}
  \label{eq:gw-imaginary-integral}
\end{equation}

For its screening response, let $x_{ps}=\omega_p-\varepsilon_s$, with $\omega_p=\varepsilon_p$ for the linearized and $\omega_p=\eqp_p$ for the implicit QP equation, and define
\begin{equation}
  \bm{\Gamma}^{\Sigma}(\nu)
  = -\frac{1}{\pi}\sum_{ps}v_p
    \frac{x_{ps}}{x_{ps}^2+\nu^2}
    \mat{b}_{ps}\mat{b}_{ps}^{\mathsf{T}}.
\end{equation}
The response is then the integral of
$\Tr[\bar{\bm{\Gamma}}^{\Sigma}(\nu)\,\mathrm{d}\bm{\Pi}(\mathrm{i}\nu)]$,
with $\bar{\bm{\Gamma}}^{\Sigma}=\mat{M}\bm{\Gamma}^{\Sigma}\mat{M}$.
Writing
\begin{equation}
  \begin{split}
    f_\nu(\Delta) &= \frac{\Delta}{\Delta^2+\nu^2},\\
    f_\nu'(\Delta) &= \frac{\nu^2-\Delta^2}{(\Delta^2+\nu^2)^2},
  \end{split}
\end{equation}
the full screening derivative at fixed $\nu$ is
\begin{equation}
  \begin{split}
    \mathrm{d}\Pi_{PQ}(\mathrm{i}\nu)
    = -4\sum_{kc}\big[&f_\nu(\Delta_{kc})\,
       \mathrm{d}(B^P_{kc}B^Q_{kc})\\
       &+B^P_{kc}B^Q_{kc}f_\nu'(\Delta_{kc})\,
       \mathrm{d}\Delta_{kc}\big].
  \end{split}
  \label{eq:gw-pi-derivative}
\end{equation}
Since $\bm{\Pi}(\mathrm{i}\nu)$ depends on neither $p$ nor $s$, it is
built and factorized once per frequency, and all pairs enter only through
$\bar{\bm{\Gamma}}^{\Sigma}(\nu)$.
Per frequency, that is, before the integration over $\nu$, the
integral-derivative term has the structure of the static screening
response with $1/\Delta$ replaced by $f_\nu(\Delta)$,
\begin{equation}
  g^P_{kc} \leftarrow -8 f_\nu(\Delta_{kc})
    \sum_Q \bar{\Gamma}^{\Sigma}_{PQ}(\nu) B^Q_{kc},
\end{equation}
supplemented by the derivative of the bare pair vectors,
$g^P_{ps} \leftarrow
-\tfrac{2}{\pi}v_p\,x_{ps}/(x_{ps}^2+\nu^2)\,
\{[\mat{M}(\mathrm{i}\nu)-\mat{1}]\mat{b}_{ps}\}_P$,
for all $p,s$.
The energy-derivative term requires $f_\nu'$ and gives the diagonal
weights
\begin{equation}
  d_c \leftarrow e_{kc}(\nu),\quad d_k \leftarrow -e_{kc}(\nu),\quad
  e_{kc}(\nu) = -4 f_\nu'(\Delta_{kc})\,
    \mat{b}_{kc}^{\mathsf{T}}\bar{\bm{\Gamma}}^{\Sigma}(\nu)\mat{b}_{kc}.
\end{equation}
These contractions are accumulated over the frequency grid as in
RI-RPA gradients.\cite{Burow2014}
The explicit dependence on the internal propagator energies gives
further diagonal weights,
\begin{equation}
  d_s^{\mathrm{I,prop}} = -\frac{1}{\pi}\int_0^\infty\!\mathrm{d}\nu\,
    \sum_p v_p W^{\mathrm{c}}_{ps,ps}(\mathrm{i}\nu)
    \frac{x_{ps}^2-\nu^2}{(x_{ps}^2+\nu^2)^2}.
\end{equation}
All derivatives with respect to the orbital energies, including the
$(v_p-w_p)$ term of equation~\eqref{eq:g0w0}, are thus collected in
diagonal weights $d_p$, and all derivatives with respect to the
three-index integrals in $g^P_{pq}$.
The orbital response follows from $\mat{g}$ and the KS canonical
conditions, as described below.

The residue term is\cite{Holzer.Klopper:Ionized.2019,Holzer:Practical.2023}
\begin{align}
  \Sigma^{\mathrm{c,R}}_{pp}(\omega)
  = & \sum_s f_s(\omega)W^{\mathrm{c}}_{ps,ps}(z_s), \\
  z_s= & |\varepsilon_s-\omega|+\mathrm{i}0^+,
  \label{eq:gw-residues}
\end{align}
where, for a Fermi energy $\varepsilon_{\mathrm{F}}$ in the gap,
\begin{equation}
  f_s(\omega)=
  \begin{cases}
    +1,&\varepsilon_{\mathrm{F}}<\varepsilon_s<\omega,\\
    -1,&\omega<\varepsilon_s<\varepsilon_{\mathrm{F}},\\
    \tfrac12-n_s,&\omega=\varepsilon_s,\\
    0,&\text{otherwise}.
  \end{cases}
\end{equation}
Here $n_s$ is one for an occupied orbital and zero for a virtual orbital.
At such coincidences, essentially occurring for every KS orbital at its own energy,
the combined integral and residue expression must be differentiated by taking
its continuous limit.
This limit is obtained by subtracting $W^{\mathrm{c}}_{ps,ps}(0)$ from the
integrand and from the residue and adding it back analytically.
Since $-\frac{1}{\pi}\int_0^\infty\mathrm{d}\nu\,x/(x^2+\nu^2)
=-\tfrac12\operatorname{sgn}(x)$ and $h_s-f_s=-\tfrac12\operatorname{sgn}(x_{ps})$
with $h_s=\tfrac12-n_s$,
\begin{equation}
  \begin{split}
    \operatorname{Re}\Sigma^{\mathrm{c}}_{pp}(\omega)
    = \sum_s\Big\{&h_s W^{\mathrm{c}}_{ps,ps}(0)
      + f_s\big[W^{\mathrm{c}}_{ps,ps}(z_s)-W^{\mathrm{c}}_{ps,ps}(0)\big]\\
     &-\frac{1}{\pi}\int_0^\infty\!\mathrm{d}\nu\,
       \big[W^{\mathrm{c}}_{ps,ps}(\mathrm{i}\nu)-W^{\mathrm{c}}_{ps,ps}(0)\big]
       \frac{x}{x^2+\nu^2}\Big\}
  \end{split}
  \label{eq:gw-subtracted}
\end{equation}
for all $\omega$, with $x=\omega-\varepsilon_s$.
Each term is continuous at $x=0$, where the residue term vanishes
because $\partial_zW^{\mathrm{c}}(0)=0$.
The second frequency derivative needed for the linearized equation is
$h_s\,\partial_z^2W^{\mathrm{c}}_{ps,ps}(0)$ at $x=0$.
The derivatives then yield
\begin{align}
  \mathrm{d}\Sigma^{\mathrm{c,R}}_{pp}
    &= \sum_s f_s\left[
       \left.\mathrm{d}W^{\mathrm{c}}_{ps,ps}\right|_{z_s}
       +\partial_z W^{\mathrm{c}}_{ps,ps}(z_s)\,\mathrm{d}z_s\right],\\
  \mathrm{d}z_s
    &= \operatorname{sgn}(\varepsilon_s-\omega)
       (\mathrm{d}\varepsilon_s-\mathrm{d}\omega).
  \label{eq:gw-residue-frequency}
\end{align}
The first term denotes a derivative at fixed frequency, including the
integral, orbital and screening response:
\begin{equation}
  \begin{split}
    \left.\mathrm{d}W^{\mathrm{c}}_{ps,ps}\right|_z
    ={}&2(\mathrm{d}\mat{b}_{ps})^{\mathsf{T}}
          (\mat{M}-\mat{1})\mat{b}_{ps}\\
       &+\mat{b}_{ps}^{\mathsf{T}}\mat{M}
          (\mathrm{d}\bm{\Pi})_z\mat{M}\mat{b}_{ps}.
  \end{split}
\end{equation}
The fixed-frequency screening derivative is obtained from
equation~\eqref{eq:gw-pi-complex}:
\begin{equation}
  \begin{split}
    (\mathrm{d}\Pi_{PQ})_z = -4\sum_{kc}\bigg[
      &\frac{\Delta_{kc}}{\Delta_{kc}^2-z^2}
       \mathrm{d}(B^P_{kc}B^Q_{kc})\\
      &-B^P_{kc}B^Q_{kc}
       \frac{\Delta_{kc}^2+z^2}{(\Delta_{kc}^2-z^2)^2}
       \mathrm{d}\Delta_{kc}\bigg].
  \end{split}
\end{equation}
The frequency derivative follows from
\begin{align}
  \partial_z W^{\mathrm{c}}_{ps,ps}
    &=\mat{b}_{ps}^{\mathsf{T}}\mat{M}
      (\partial_z\bm{\Pi})\mat{M}\mat{b}_{ps},\\
  \partial_z\Pi_{PQ}(z)
    &=-8z\sum_{kc}\frac{B^P_{kc}B^Q_{kc}\Delta_{kc}}
                          {(\Delta_{kc}^2-z^2)^2}.
\end{align}
For $\partial_\xi\sigma_p$ in equation~\eqref{eq:g0w0},
$\mathrm{d}\omega=0$ in equation~\eqref{eq:gw-residue-frequency}, and the external
frequency derivatives of both the integral and the residues enter $Z_p$.
Only the real part contributes to $\sigma_p$.

Since $\Lambda$ depends on the diagonal of the self-energy, it is not
invariant under rotations within the occupied or virtual space.
The corresponding multipliers include the external-orbital contributions
from $\operatorname{Re}\Sigma_{pq}(\eqp_p)-v^{\mathrm{xc}}_{pq}$,
as well as the orbital response inside $G_0$ and $W_0$.
Both are contained in
\begin{equation}
  Y_{rp} = \sum_{Pq} B^P_{rq}\left(g^P_{pq}+g^P_{qp}\right)
         - 2v_p\,v^{\mathrm{xc}}_{rp},
\end{equation}
where $\mat{g}$ now also contains the exchange and, for hybrid starting
points, the nonlocal part of $v^{\mathrm{xc}}$, and the last term is the
rotation of the semilocal $v^{\mathrm{xc}}_{pp}$.
Occupied--virtual rotations enter the right-hand side of the Z-vector
equation as $Y_{ai}-Y_{ia}$, and the overlap term as $\tfrac12Y_{ia}$.
Rotations within the occupied or the virtual space are eliminated with
the canonical conditions of the KS reference,
$U^{\xi}_{rp}=(F^{\xi}_{rp}-\varepsilon_pS^{\xi}_{rp})/(\varepsilon_p-\varepsilon_r)$
for $r\neq p$ and $U^{\xi}_{pp}=-\tfrac12S^{\xi}_{pp}$, together with
$\mathrm{d}\varepsilon_p=F^{\xi}_{pp}-\varepsilon_pS^{\xi}_{pp}$.
This gives the response density
\begin{equation}
  D^{\Sigma}_{rp} = \delta_{rp}\,d_p
    + (1-\delta_{rp})\,\frac{Y_{rp}-Y_{pr}}{2(\varepsilon_p-\varepsilon_r)},
\end{equation}
which is contracted with $F^{\xi}_{rp}$ in the same way as $\mat{T}$,
and the energy-weighted density
\begin{equation}
  \Theta^{\Sigma}_{rp} = \tfrac14\left(Y_{rp}+Y_{pr}\right)
    + \tfrac12 D^{\Sigma}_{rp}\left(\varepsilon_r+\varepsilon_p\right).
\end{equation}
This treatment applies unchanged to the external orbital $p$ and to ev$GW$
(Section~\ref{sec:qp_evgw}).
Because of the energy denominators, the antisymmetric part of $\mat{Y}$
must vanish within degenerate KS levels, which holds for symmetry-related
partners but not for accidental degeneracies.
The semilocal part of $v^{\mathrm{xc}}_{pp}$ further requires the
contraction of the XC kernel with $\sum_pv_p|\varphi_p\rangle\langle\varphi_p|$
in the Z-vector equation, in $\bm{\Theta}$ and in the explicit derivative.

The screening is factorized once per frequency and shared by all orbital pairs.
For a few corrected orbitals, the number of residue terms is at most of the order of the
number of imaginary frequencies, each costing as much as one frequency of an
RI-RPA gradient, so the correction terms keep the formal $\mathcal{O}(N^4)$ scaling.
An operation count estimates their cost at about three to four times
that of the preceding $G_0W_0$ calculation.

\subsection{Gradient correction for RI-CD-ev$GW$}
\label{sec:qp_evgw}
In ev$GW$,\cite{Faleev2004,vanSchilfgaarde2006,Kotani2007,Shishkin2007,Blase2011,Caruso2012,Kaplan2016} the QP energies also enter the updated
$G$ and $W$ in each iteration, without updating the KS orbitals.
I define $\mat{E}=(\eqp_p)_p$ and $\sigma_p(\omega,\mat{E};\xi)$ 
to distinguish the external frequency from these internal KS energies.
At a converged ev$GW$ QP solution,
\begin{equation}
  E_p = \varepsilon_p+\sigma_p(E_p,\mat{E};\xi).
\end{equation}
Its Jacobian is
\begin{equation}
  J_{pq} =
  \left[\delta_{pq}\,\partial_\omega\sigma_p(\omega,\mat{E};\xi)
       +\left.\partial_{E_q}\sigma_p(\omega,\mat{E};\xi)
         \right|_{\omega}\right]_{\omega=E_p}.
  \label{eq:evgw-jacobian}
\end{equation}
The second derivative holds $\omega$ fixed and therefore excludes the
external-frequency contribution already contained in the first term.
Differentiation yields
\begin{equation}
  \begin{split}
    (\mat{1}-\mat{J})\frac{\mathrm{d}\mat{E}}{\mathrm{d}\xi}
    &= \frac{\mathrm{d}\bm{\varepsilon}}{\mathrm{d}\xi}+\mat{a},\\
    a_p &= \left.\partial_\xi\sigma_p(\omega,\mat{E};\xi)
          \right|_{\omega=E_p,\,\mat{E}\ \mathrm{fixed}},
  \end{split}
\end{equation}
where the partial derivative also holds $\omega$ fixed.
The weighted correction follows from the adjoint equation
\begin{align}
  (\mat{1}-\mat{J})^{\mathsf{T}}\mat{v} &= \mat{w},\\
  \sum_p w_p\frac{\mathrm{d}c_p}{\mathrm{d}\xi}
    &= \mat{v}^{\mathsf{T}}\mat{a}
      +\sum_p(v_p-w_p)\frac{\mathrm{d}\varepsilon_p}{\mathrm{d}\xi}.
  \label{eq:evgw-gradient}
\end{align}
In the integral and residue formulas, the internal $\varepsilon_s$ and
$\Delta_{kc}$ are replaced by $E_s$ and $E_c-E_k$.
Their derivatives with respect to $\mat{E}$, including the moving residue
frequencies, enter $\mat{J}$. $\mat{a}$ collects the remaining geometric
dependence through KS orbitals, Coulomb, exchange and $v^{\mathrm{xc}}$
as usual.\cite{Furche2002}
For $s=p$, the external and internal energies coincide, and the continuous
combined-expression prescription above applies to the Jacobian as well.
Thus, the internal-energy weights belong to the QP adjoint equation rather
than being counted again as KS-energy derivatives in $\mat{a}$.
For a few corrected orbitals, $\mat{J}$ is small. Its rows follow from the
internal-energy weights of each $\sigma_p$, evaluated with the same factorized
screening, and the adjoint equation is solved directly.
For many corrected orbitals, products $\mat{J}^{\mathsf{T}}\mat{v}$ can instead be
evaluated by differentiating
$\sum_pv_p\sigma_p(E_p,\mat{E};\xi)$ with respect to $\mat{E}$ at fixed
$\mat{v}$ and fixed KS quantities, allowing an iterative adjoint solution.
For $G_0W_0$, the internal QP-energy dependence is absent,
$\mat{J}$ is diagonal and $v_p=Z_pw_p$ ($Z_p^{\mathrm{lin}}$ for the linearized equation).

\subsection{Quasiparticle energies with fixed corrections}
\label{sec:qp}
QP energies are introduced as $\eqp_p = \varepsilon_p + c_p$, where the $GW$
corrections $c_p$ are kept fixed.
Since the BSE functional depends on the diagonal orbital energies only, the
contributions of rotations among occupied or among virtual orbitals have
to be considered explicitly.
They are obtained by writing the QP Hamiltonian in a rotationally invariant
form, $\hat{H}^{\mathrm{QP}} = \hat{F} + \sum_p c_p|\varphi_p\rangle\langle\varphi_p|$,
with the canonical KS orbitals $\varphi_p$.
Re-canonicalization of the perturbed orbitals rotates the projectors.
In a fixed orthonormal frame, the non-canonical orbital response
$U^{\xi}_{pq} = F^{\xi}_{pq}/(\varepsilon_q-\varepsilon_p)$ gives
\begin{equation}
  \begin{split}
    & \big(H^{\mathrm{QP}}\big)^{\xi}_{pq} = r_{pq}\,F^{\xi}_{pq},\\
    &r_{pq} = \frac{\eqp_q-\eqp_p}{\varepsilon_q-\varepsilon_p}, \qquad r_{pp} = 1,
  \end{split}
\end{equation}
for $p,q$ both occupied or both virtual.
Hence, the off-diagonal elements of $\mat{T}$ are scaled by $r_{pq}$
computed with the energies used in $\Abse$, and those of
$\mat{D}^{\Pi}$ by $r_{pq}$ computed with the energies used in $\bm{\Pi}$.
For degenerate KS orbitals, $r_{pq}=1$, which presumes equal corrections
within a degenerate set.
The same factor was introduced as an approximation for
electron--phonon matrix elements,\cite{IsmailBeigi2003,DelGrande2025} but here
it is exact for geometry-independent corrections.
For a geometry-dependent AO overlap, let $F^{\xi}_{pq}$ and
$S^{\xi}_{pq}$ denote the matrix derivatives transformed with the
unperturbed MO coefficients.
In the symmetrically orthonormalized frame, the corresponding expression is
\begin{equation}
  \big(\widetilde H^{\mathrm{QP}}\big)^{\xi}_{pq}
  = r_{pq}\left[F^{\xi}_{pq}
    -\frac{\varepsilon_p+\varepsilon_q}{2}S^{\xi}_{pq}\right].
  \label{eq:qp-metric-derivative}
\end{equation}
With the gradient convention $-\sum_{pq}\Theta_{pq}S^{\xi}_{pq}$, the
QP-energy part of the TDDFT energy-weighted density therefore requires
the addition
\begin{equation}
  \Delta\Theta^T_{pq} = \tfrac12 T_{pq}
  \left[r_{pq}(\varepsilon_p+\varepsilon_q)
        -(\eqp_p+\eqp_q)\right].
\end{equation}
Here $T_{pq}$ denotes the density before scaling by $r_{pq}$.
The screening density must be treated consistently: its overlap
coefficient is
$\tfrac12 r^{\Pi}_{pq}(\varepsilon_p+\varepsilon_q)D^{\Pi}_{pq}$,
with $r^{\Pi}_{pq}$ formed from the energies used in $\bm{\Pi}$.
If its energy-weighted contribution was first constructed with QP
energies, the analogous replacement is required.
The omitted geometry dependence of $c_p$ is weighted by both the
excitation and screening densities as described above.

%%%%%%%%%%%%%%%%%%%%%%%%%%%%%%%%%%%%%%%%%%%%%%%%%%%%%%%%%%%%%%%%%%%%%
\section{Computational Details}
\label{sec:compdetails}
%%%%%%%%%%%%%%%%%%%%%%%%%%%%%%%%%%%%%%%%%%%%%%%%%%%%%%%%%%%%%%%%%%%%%
The gradients were assessed through excited-state geometry optimizations for the
0--0 benchmark set of Loos and Jacquemin,\cite{Loos2019} which comprises
69 singlet and 31 triplet excited states of 71 closed-shell molecules.
Optimizations were run for all 100 states with BHLYP and PBE0 references and
with $G_0W_0$ and ev$GW$ quasiparticle energies.
For comparison, TD-DFT optimizations with the same reference functionals 
were performed. The reference excited-state minima were taken from
the supporting information of Ref.~\citenum{Loos2019} and were optimized
with EOM-CCSD (singlets) and U-CCSD (triplets) in the def2-TZVPP basis.
The radical set of that benchmark was not considered.

All BSE gradient calculations were carried out with a development version of the
 \textsc{egrad} module of \textsc{Turbomole},
\cite{Balasubramani.Chen.ea:TURBOMOLE.2020,Franzke.Holzer.ea:TURBOMOLE.2023} 
without point-group symmetry, using the def2-TZVPP basis
set\cite{Weigend2005} together with the corresponding density (KS ground state) or
MP2-optimized ($GW$, BSE) auxiliary basis sets
for the RI approximation.\cite{Eichkorn1995,Weigend.Haser.ea:RI-MP2.1998}
Reference determinants were obtained with the global hybrid functionals
BHLYP\cite{Becke1993,Lee1988} (50\,\% exact exchange) and
PBE0 (25\,\% exact exchange).\cite{Perdew1996,Adamo1999}
$G_0W_0$ and ev$GW$ quasiparticle energies were computed with the RI implementation\cite{Ren2012,Holzer.Klopper:Ionized.2019}
and contour deformation\cite{Holzer.Klopper:Ionized.2019,Holzer:Practical.2023}
using 128 imaginary-frequency points at every geometry of the optimization.
The highest 10 occupied and lowest 10 unoccupied orbitals have been
corrected, while the remaining orbitals were shifted accordingly with the
quasiparticle correction of the closest corrected state.
SCF energies were converged to $10^{-8}\,E_\mathrm{h}$, and the DFT integration grid
was a Treutler--Ahlrichs grid of size~3.\cite{Treutler.Ahlrichs:Efficient.1995,Treutler:Entwicklung.1995}
The TD-DFT calculations used the same orbital and auxiliary basis sets, with the
auxiliary basis for the response chosen as in the BSE runs, and the
seminumerical exchange approximation\cite{Holzer:improved.2020} with a \textit{veryfine} grid.
The optimizations used a the new \textsc{aiopt} module of
\textsc{Turbomole}\cite{Franzke.Sharma.ea:Recent.2026}
and were converged to an energy change of $10^{-6}\,E_\mathrm{h}$.
For the comparison with the full quasiparticle gradient (Section~\ref{sec:geometry}), 
the optimizations of the 62 states of the common subset were repeated with the full 
quasiparticle gradient for both functionals and both $GW$ variants, 
with otherwise identical settings. The charge-transfer state of 
4-(Dimethylamino)benzonitril (DMABN) (Section~\ref{sec:dmabn}) was optimized with 
a PBE0/def2-TZVP reference, the full quasiparticle gradient and both $GW$ variants, 
starting from the EOM-CCSD structure of Ref.~\citenum{Knysh2022}.

%%%%%%%%%%%%%%%%%%%%%%%%%%%%%%%%%%%%%%%%%%%%%%%%%%%%%%%%%%%%%%%%%%%%%
\section{Results}
\label{sec:results}
%%%%%%%%%%%%%%%%%%%%%%%%%%%%%%%%%%%%%%%%%%%%%%%%%%%%%%%%%%%%%%%%%%%%%
\subsection{Success of the optimizations}
\label{sec:success}
Within the fixed quasiparticle approach outlined in Section~\ref{sec:qp}, 
82 (BHLYP) and 65 (PBE0) of the 100 states were successfully optimized with $G_0W_0$, and 
90 (BHLYP) and 86 (PBE0) with ev$GW$. Successful means that the optimization converged to a
structure with the bond graph and bond angles of the reference, which excludes triplet thioacetaldehyde 
(the reference is a different isomer) and, for BHLYP, isocyanogen (which stays linear).
Only 62 states worked with every combination of KS reference and $GW$ method.
This common subset was subsequently used 
to perform optimizations with the full quasiparticle gradients outlined 
in Sections~\ref{sec:qp_g0w0} and \ref{sec:qp_evgw}.
Most failed optimizations can be traced back to triplet instabilities, 
which are severe within $GW$-BSE and stem from the well-known
underestimation of triplet excitation energies.\cite{Gui.Holzer.ea:Accuracy.2018}
The remaining failures are consistent with the BSE Hessian 
not being guaranteed to be positive definite for a KS reference, 
an issue that is most pronounced for $G_0W_0$-based optimizations.
ev$GW$ generally increases the gap between occupied and unoccupied
orbitals, which explains the significant decrease in unstable optimizations.
Likewise, the larger amount of exact exchange in BHLYP
leads to an increased gap between the occupied and virtual states, 
making a positive definite BSE Hessian more likely. 
Additionally, the solution of the quasiparticle equations
within $G_0W_0$, done by linearization in this work (Section~\ref{sec:qp_g0w0}), 
is prone to yield renormalization factors $Z_p$ far from unity, 
in particular small ones, close to a pole of the self-energy.
The quasiparticle corrections then change erratically between neighboring geometries,
so that the energy varies irregularly, which can lead to diverging optimizations. 
This is remedied by using ev$GW$.
The TD-DFT optimizations draw a different picture. For the 86 states where at least the ev$GW$
approach succeeded, only 74 states were successfully optimized with 
BHLYP, while 84 optimizations finished with PBE0.
All 14 failed TD-DFT optimizations involved triplet states, 
showing that TD-DFT is also troubled by triplet instabilities, 
and that the failures are not exclusive to the $GW$-BSE method.
Indeed, ev$GW$-BSE proves to be a rather robust method already for 
this set of small molecules, while TD-DFT is widely and successfully used to optimize 
excited-state geometries despite its known deficiencies concerning
triplet instabilities. Optimization of triplet states with either
TD-DFT or the $GW$-BSE method therefore needs to be treated carefully.

\subsection{Analysis of $GW$-BSE geometries}
\label{sec:geometry}

Table~\ref{tab:errors} gives the deviations of the optimized PBE0-based geometries 
for BSE@$G_0W_0$ and BSE@ev$GW$, each with fixed quasiparticle corrections and with 
the full quasiparticle gradient, and for TD-DFT. All entries refer to the same states, 
56 states of the common subset for which, in addition, the optimizations with the PBE0 
reference were successful with the fixed and full quasiparticle gradient and with TD-DFT, 
i.e., with all five approaches. Table~\ref{tab:pairs} analogously 
contains the 59 states of the common subset whose BSE@ev$GW$ optimizations with fixed
corrections and with the full gradient succeeded with both functionals, and Table~\ref{tab:perstate} lists all 62 states of the 
common subset, outlining which optimizations were unsuccessful.

\begin{table*}[htb]
	\caption{Deviations of optimized geometries with the PBE0 reference for BSE@$G_0W_0$ and BSE@ev$GW$ with fixed quasiparticle corrections
		(fixed) and with the full quasiparticle gradient (full), and for TD-DFT,
		for the common subset of the 45 singlets and 11 triplets whose
		optimizations were successful with all five approaches:
		mean signed deviation (MSD), mean absolute deviation (MAD) and root-mean-square deviation (RMSD)
		relative to the EOM-CCSD (singlets) and U-CCSD (triplets) structures,
		for bond lengths (\AA) and bond angles at non-hydrogen atoms (degrees).
		$N$ is the number of bonds or angles.}
	\label{tab:errors}
	\footnotesize
	\setlength{\tabcolsep}{2.5pt}
	\begin{tabular}{llrrrrrrrrrrrrrrrrr}
		\toprule
		& & & \multicolumn{6}{c}{BSE@$G_0W_0$} & \multicolumn{6}{c}{BSE@ev$GW$} & \multicolumn{3}{c}{TD-DFT} \\
		\cmidrule(lr){4-9}\cmidrule(lr){10-15}\cmidrule(lr){16-18}
		& & & \multicolumn{3}{c}{fixed} & \multicolumn{3}{c}{full} & \multicolumn{3}{c}{fixed} & \multicolumn{3}{c}{full} & & & \\
		\cmidrule(lr){4-6}\cmidrule(lr){7-9}\cmidrule(lr){10-12}\cmidrule(lr){13-15}
		& & $N$ & MSD & MAD & RMSD & MSD & MAD & RMSD & MSD & MAD & RMSD & MSD & MAD & RMSD & MSD & MAD & RMSD \\
		\midrule
		singlets & heavy-atom bonds & 176 & $-$0.008 & 0.010 & 0.015 & $-$0.002 & 0.009 & 0.015 & $-$0.009 & 0.011 & 0.015 & $-$0.003 & 0.009 & 0.015 & $-$0.005 & 0.011 & 0.016 \\
		& X--H bonds & 121 & $-$0.001 & 0.003 & 0.005 & $+$0.005 & 0.005 & 0.006 & $-$0.001 & 0.003 & 0.005 & $+$0.005 & 0.005 & 0.006 & $+$0.005 & 0.005 & 0.006 \\
		& angles & 389 & 0.00 & 0.50 & 0.89 & $+$0.13 & 0.64 & 1.24 & $+$0.06 & 0.49 & 0.85 & $+$0.05 & 0.57 & 0.99 & $+$0.09 & 0.56 & 0.97 \\
		\midrule
		triplets & heavy-atom bonds & 48 & $-$0.008 & 0.020 & 0.027 & 0.000 & 0.013 & 0.022 & $-$0.007 & 0.020 & 0.029 & 0.000 & 0.020 & 0.027 & $-$0.005 & 0.013 & 0.022 \\
		& X--H bonds & 36 & $-$0.001 & 0.002 & 0.004 & $+$0.005 & 0.005 & 0.006 & 0.000 & 0.002 & 0.004 & $+$0.005 & 0.005 & 0.006 & $+$0.005 & 0.006 & 0.006 \\
		& angles & 116 & $+$0.21 & 0.96 & 1.81 & $+$0.16 & 0.96 & 1.78 & $+$0.14 & 1.10 & 1.97 & $+$0.19 & 1.21 & 2.02 & $+$0.25 & 0.93 & 1.77 \\
		\bottomrule
	\end{tabular}
\end{table*}

With fixed corrections, the heavy-atom bond lengths of the singlet states are on average too short with all methods,
and BSE and TD-DFT are of the same quality.
For the triplets, TD-DFT is closer to the reference than BSE.
The bonds to hydrogen show the opposite trend: BSE reproduces them almost exactly, whereas TD-DFT overestimates them.
The full quasiparticle gradient lengthens the heavy-atom bonds and removes most of their systematic underestimation.
For the singlets, BSE@ev$GW$ is then closer to the reference than TD-DFT for most states.
For the triplets, BSE@$G_0W_0$ reaches the quality of TD-DFT, whereas BSE@ev$GW$ does not improve.
The price is the X--H bonds, which are then overestimated as much as with TD-DFT, so that the advantage of BSE
for these bonds disappears, and a slight increase of the bond-angle deviations.
BSE with PBE0 and fixed corrections therefore does not improve the heavy-atom bond lengths relative to TD-DFT
with the same functional, but avoids its overestimation of the X--H bonds, whereas
with the full gradient the heavy-atom bonds improve at the expense of the X--H bonds.

Table~\ref{tab:pairs} groups the heavy-atom bond deviations of BSE@ev$GW$ by element pair.
With fixed corrections, the bonds are too short for all element pairs, most strongly for bonds to oxygen and sulfur,
where the BHLYP-based deviations are larger than those of PBE0.
The full gradient reduces the deviations of these polar bonds most.
The EOM-CCSD and U-CCSD reference structures are themselves approximate\cite{Budzak2017}:
CCSD is known to give overly localized excited-state structures, with, for example, a
mean error of $-0.021$~\AA\ for excited-state C=O bond lengths,\cite{Loos2019} 
so that the systematic shortening cannot be assigned to the BSE gradients alone.
The benchmark comparison is also limited to geometries: no energies were compared,
and the statistics include only states for which the optimizations being compared
were successful, which may favor well-behaved cases.
Deviations for individual states are listed in Table~\ref{tab:perstate}.

Earlier work on BSE excited-state geometries used much smaller test sets.
Çaylak and Baumeier optimized four molecules with numerical gradients and found mean relative errors
of about 1\,\% for $G_0W_0$ and 1.4--1.5\,\% for ev$GW$ with respect to CASPT2 for a PBE0 starting point,
smaller than those of CC2 and TD-DFT.\cite{Caylak2021}
The mean relative deviation of the heavy-atom bond lengths in the present set is 0.9\,\% for both
BSE variants with fixed corrections, 0.7\,\% (BSE@$G_0W_0$) and 0.8\,\% (BSE@ev$GW$) with the full gradient,
and 0.8\,\% for TD-DFT (PBE0, common subset of Table~\ref{tab:errors}), which is of the same size.
T\"olle et al.\ reported analytic BSE@$G_0W_0$ gradients with a HF starting point and compared the changes
of selected bond lengths upon excitation with CC3 and CCSDR(3) results.\cite{Toelle2025BSE}
For formaldehyde, thio- and selenoformaldehyde, acetaldehyde, acrolein, glyoxal and acetylene,
which are also part of the present set, their deviations lie between $-0.005$ and $+0.020$~\AA\ for
the C=X bonds.
For the same bonds, the PBE0-based BSE structures here (fixed corrections) deviate from the EOM-CCSD references by
$-0.031$ to $-0.001$~\AA, and the TD-DFT structures by $-0.031$ to $+0.009$~\AA.
The two comparisons use different references and starting points and are therefore only qualitative.
They indicate that the starting point has an influence on the sign of the deviation.

\begin{table}[htb]
\caption{Deviations (\AA) of heavy-atom bond lengths by element pair for BSE@ev$GW$,
  singlets and triplets together ($N$ bonds), with fixed quasiparticle corrections (fixed) and with the full quasiparticle gradient (full),
  for the 59 states of the common subset with successful optimizations in both cases and with both functionals.
  Only pairs with $N\ge 5$ are listed.}
\label{tab:pairs}
\begin{tabular}{lrrrrr}
\toprule
 & & \multicolumn{2}{c}{BHLYP} & \multicolumn{2}{c}{PBE0} \\
pair & $N$ & MSD & MAD & MSD & MAD \\
\midrule
\multicolumn{6}{c}{fixed QP correction} \\
C--C & 136 & $-$0.008 & 0.016 & $-$0.006 & 0.011 \\
C--N & 34 & $-$0.015 & 0.015 & $-$0.007 & 0.007 \\
C--O & 33 & $-$0.033 & 0.033 & $-$0.016 & 0.018 \\
C--F & 12 & $-$0.010 & 0.010 & $-$0.001 & 0.002 \\
C--S & 7 & $-$0.036 & 0.036 & $-$0.031 & 0.031 \\
\midrule
\multicolumn{6}{c}{full QP correction} \\
C--C & 136 & $-$0.009 & 0.014 & $-$0.004 & 0.011 \\
C--N & 34 & $-$0.012 & 0.012 & $+$0.002 & 0.005 \\
C--O & 33 & $-$0.015 & 0.017 & $-$0.003 & 0.012 \\
C--F & 12 & $-$0.009 & 0.009 & $-$0.001 & 0.003 \\
C--S & 7 & $-$0.016 & 0.016 & $-$0.009 & 0.009 \\
\bottomrule
\end{tabular}
\end{table}

With fixed corrections, the BHLYP-based structures deviate more 
from the reference than the PBE0-based structures for all three methods, 
and PBE0 is closer to the reference for the large majority of the states.
The full quasiparticle gradient lengthens the bonds of the BHLYP-based 
structures most and thereby reduces, but does not remove, the functional dependence
(Table~\ref{tab:pairs}), in line with the reduced starting point 
dependence of $GW$.\cite{Bruneval.Marques:Benchmarking.2013} 
However, especially the X--H bonds with PBE0 become too long if 
the full quasiparticle correction is used. The element-pair deviations 
of BSE@$G_0W_0$ are similar to those of ev$GW$.
Eigenvalue self-consistency therefore has a negligible influence on the optimized structures, 
the per-state heavy-atom deviations of BSE@ev$GW$ and BSE@$G_0W_0$ are nearly 
identical for both functionals. Its main effect in these calculations is an 
increase in the number of successful optimizations, and the added
robustness is a key feature of ev$GW$-BSE during geometry optimizations.
Self-consistency therefore does not reduce the dependence of the geometries on the
KS reference, which is in stark contrast to the excitation energies.
For the latter, it is routinely found that ev$GW$ reduces the functional dependence
considerably.\cite{Korbel.Boulanger.ea:Benchmark.2014,Gui.Holzer.ea:Accuracy.2018}

\begin{table*}[htb]
\caption{Mean absolute deviation (\AA) of the heavy-atom bond lengths for each state of the common subset
  of 62 states (46 singlet and 16 triplet states of 48 molecules), for BSE@$G_0W_0$ and BSE@ev$GW$ with the full quasiparticle gradient and for TD-DFT.
  The 62 states are those for which the optimizations with fixed quasiparticle corrections succeeded for both
  functionals and both $GW$ variants; TD-DFT and the full quasiparticle gradient were not part of this criterion.
  ``(T)'' marks triplets, ``(2)'' the second state of a symmetry, and -- an unsuccessful optimization
  with the full quasiparticle gradient or with TD-DFT.}
\label{tab:perstate}
\footnotesize
\setlength{\tabcolsep}{3pt}
\begin{tabular}{lrrrrrrlrrrrrr}
\toprule
 & \multicolumn{2}{c}{$G_0W_0$} & \multicolumn{2}{c}{ev$GW$} & \multicolumn{2}{c}{TD-DFT} & & \multicolumn{2}{c}{$G_0W_0$} & \multicolumn{2}{c}{ev$GW$} & \multicolumn{2}{c}{TD-DFT} \\
state & BHLYP & PBE0 & BHLYP & PBE0 & BHLYP & PBE0 & state & BHLYP & PBE0 & BHLYP & PBE0 & BHLYP & PBE0 \\
\midrule
Acetaldehyde & 0.012 & 0.018 & 0.012 & 0.006 & 0.017 & 0.011 & 4\emph{H}-Pyran-4-one & 0.011 & 0.011 & 0.011 & 0.007 & 0.011 & 0.009 \\
Acetaldehyde (T) & 0.020 & 0.014 & 0.021 & 0.014 & 0.024 & 0.017 & 4\emph{H}-Pyran-4-one (T) & 0.018 & 0.003 & 0.014 & 0.018 & -- & 0.006 \\
Acetone & 0.015 & 0.021 & 0.014 & 0.012 & 0.015 & 0.013 & H$_2$C$_3$ & 0.016 & 0.004 & 0.015 & 0.003 & 0.017 & 0.006 \\
Acetyl cyanide & 0.011 & 0.009 & 0.011 & 0.004 & 0.015 & 0.013 & HCN & 0.007 & 0.006 & 0.013 & 0.006 & 0.007 & 0.010 \\
Acetylene & 0.011 & 0.002 & 0.009 & 0.008 & 0.005 & 0.009 & HCP & 0.034 & 0.030 & 0.013 & 0.009 & 0.008 & 0.011 \\
Acetylene (2) & 0.010 & 0.004 & 0.008 & 0.008 & 0.002 & 0.008 & HPO & 0.022 & 0.019 & 0.027 & 0.004 & 0.036 & 0.013 \\
Acrolein & 0.010 & 0.007 & 0.011 & 0.008 & 0.017 & 0.025 & HPS & 0.001 & 0.004 & 0.011 & 0.000 & 0.030 & 0.015 \\
Acrolein (T) & 0.023 & 0.008 & 0.024 & 0.021 & -- & 0.018 & HSiF & 0.006 & 0.006 & 0.005 & 0.006 & 0.008 & 0.006 \\
Aniline & 0.010 & 0.004 & 0.010 & 0.004 & 0.012 & 0.006 & Oxalyl chloride (T) & 0.045 & 0.039 & 0.046 & 0.040 & 0.039 & 0.038 \\
Benzaldehyde (T) & 0.038 & 0.010 & 0.040 & 0.034 & -- & 0.005 & Oxalyl fluoride & 0.010 & 0.003 & 0.010 & 0.005 & 0.011 & 0.003 \\
Benzonitrile & 0.011 & 0.006 & 0.011 & 0.006 & 0.012 & 0.007 & \emph{p}-Dicyanobenzene & 0.030 & 0.006 & 0.011 & 0.006 & 0.028 & 0.029 \\
Benzoquinone & 0.018 & -- & 0.018 & 0.015 & 0.019 & 0.016 & \emph{p}-Diethynylbenzene & 0.027 & 0.033 & -- & 0.035 & 0.026 & 0.027 \\
Benzoquinone (T) & 0.007 & -- & 0.007 & 0.004 & 0.008 & 0.002 & \emph{p}-Difluorobenzene & 0.010 & 0.002 & 0.010 & 0.003 & 0.012 & 0.005 \\
CF$_2$ & 0.004 & 0.001 & 0.007 & 0.001 & 0.015 & 0.007 & Phenylacetylene & 0.033 & 0.035 & 0.034 & 0.037 & 0.034 & 0.031 \\
Cyanoacetylene & 0.012 & 0.005 & 0.015 & 0.017 & 0.010 & 0.011 & Propynal (T) & 0.017 & 0.002 & 0.017 & 0.009 & -- & 0.012 \\
Cyanoacetylene (2) & 0.011 & 0.008 & 0.011 & 0.012 & 0.008 & 0.005 & Pyrazine & 0.010 & 0.002 & 0.010 & 0.002 & 0.010 & 0.004 \\
Cyanogen & 0.013 & 0.008 & 0.013 & 0.013 & 0.008 & 0.006 & Pyrazine (T) & 0.009 & 0.001 & 0.009 & 0.002 & 0.008 & 0.003 \\
Cyanogen (T) & 0.010 & -- & 0.011 & 0.017 & -- & -- & Pyrimidine & 0.010 & 0.002 & 0.009 & 0.003 & 0.011 & 0.004 \\
Cyanogen (2) & 0.012 & 0.007 & 0.013 & 0.012 & 0.009 & 0.004 & Pyrimidine (T) & 0.011 & 0.003 & 0.011 & 0.003 & 0.011 & 0.005 \\
2-Cyclopenten-1-one & 0.011 & 0.013 & 0.011 & 0.006 & 0.013 & 0.016 & Selenoformaldehyde & 0.002 & 0.014 & 0.003 & 0.002 & 0.026 & 0.011 \\
2-Cyclopenten-1-one (T) & 0.016 & -- & 0.016 & -- & -- & 0.013 & Silylidene & 0.013 & 0.022 & 0.021 & 0.006 & 0.022 & 0.012 \\
4-Cyclopentene-1,3-dione (T) & 0.012 & -- & 0.011 & 0.009 & -- & 0.006 & SO$_2$ (T) & 0.040 & 0.067 & 0.040 & 0.063 & 0.041 & 0.058 \\
Diacetylene & 0.012 & 0.008 & 0.012 & 0.015 & 0.008 & 0.007 & Tetrazine & -- & 0.002 & -- & 0.002 & 0.062 & 0.004 \\
Diacetylene (2) & 0.010 & 0.006 & 0.010 & 0.011 & 0.007 & 0.005 & Thioacetaldehyde & 0.009 & 0.011 & 0.011 & 0.008 & 0.021 & 0.013 \\
Difluorodiazirine & 0.011 & 0.005 & 0.010 & 0.005 & 0.009 & 0.002 & Thioacetone & 0.008 & 0.011 & 0.010 & 0.008 & 0.018 & 0.013 \\
2,6-Difluoropyridine & 0.010 & 0.001 & 0.010 & 0.002 & 0.011 & 0.003 & Thioacetone (T) & 0.016 & 0.013 & 0.018 & 0.016 & 0.025 & 0.020 \\
Fluorobenzene & 0.010 & 0.003 & 0.010 & 0.003 & 0.012 & 0.005 & Thiocarbonyl bromide & 0.010 & 0.004 & 0.010 & 0.010 & 0.018 & 0.012 \\
Formaldehyde & 0.004 & 0.001 & 0.016 & 0.014 & 0.033 & 0.017 & Thioformaldehyde & 0.010 & 0.014 & 0.013 & 0.007 & 0.033 & 0.016 \\
Formyl fluoride & 0.015 & 0.010 & 0.015 & 0.006 & 0.018 & 0.008 & Thiophosgene & 0.006 & 0.023 & 0.009 & 0.002 & 0.021 & 0.009 \\
Glyoxal & 0.011 & 0.016 & 0.011 & 0.004 & 0.014 & 0.008 & Thiopropynal (T) & 0.014 & -- & 0.016 & 0.008 & -- & 0.012 \\
Glyoxal (T) & 0.006 & 0.017 & 0.007 & 0.007 & 0.010 & 0.007 & \emph{trans}-Hydroquinone & 0.010 & 0.003 & 0.010 & 0.004 & 0.012 & 0.005 \\
\bottomrule
\end{tabular}
\end{table*}

\subsection{Fixed quasiparticle corrections}
\label{sec:fixedqp}

\begin{figure}[htb!]
	\centering
	\includegraphics[width=0.8\columnwidth]{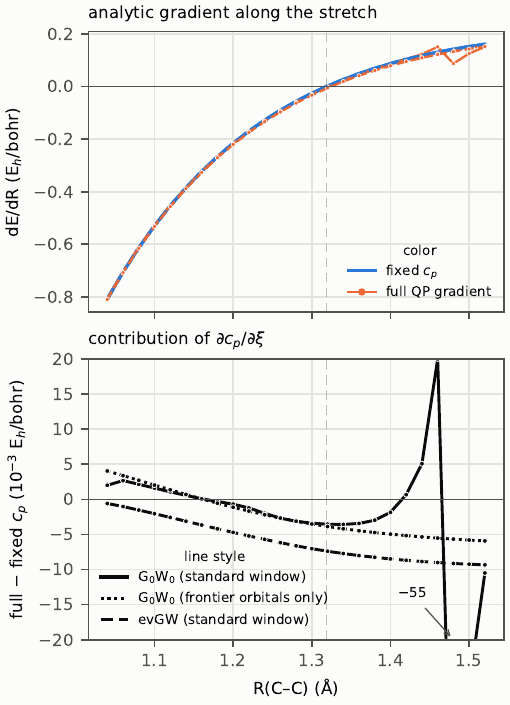}
	\caption{Analytic gradient of the $1\,\Sigma_u^-$ state of linear 
		acetylene along the symmetric C--C stretch (PBE0).
		Top: gradient with fixed (blue) and with the full 
		quasiparticle gradient (orange) for $G_0W_0$ with the standard window, 
		i.e., all occupied and the lowest ten virtual orbitals
		corrected (solid), 
		$G_0W_0$ with a window containing only the degenerate HOMO and LUMO pairs (dotted),
		and ev$GW$ with the standard window (dashed).
		Bottom: difference between the gradients with the full and 
		with the fixed quasiparticle corrections, the contribution 
		of $\partial c_p/\partial\xi$ (same line styles, clipped).
		The vertical dashed line marks the minimum along the stretch, 
		the zero of the gradient with fixed corrections at 1.319~\AA.}
	\label{fig:fixedqp}
\end{figure}

The effect of neglecting $\partial c_p/\partial\xi$ was assessed for the
$1\,\Sigma_u^-$ state of linear acetylene along the symmetric C--C stretch
with fixed C--H distances (PBE0). The results are shown in 
Figure~\ref{fig:fixedqp}.

At 25 C--C distances between 1.04 and 1.52~\AA\ 
(step 0.02~\AA), the gradient was obtained analytically with fixed quasiparticle 
corrections as outlined in Section~\ref{sec:qp}, and with the full quasiparticle 
response of Sections~\ref{sec:qp_g0w0} and \ref{sec:qp_evgw}, which includes 
$\partial c_p/\partial\xi$. The $1\,\Sigma_u^-$ state is the lowest non-degenerate 
root and was selected as such at every geometry. Three variants were considered: 
$G_0W_0$ and ev$GW$ with the standard window, i.e., all seven occupied and the lowest 
ten virtual orbitals corrected, and $G_0W_0$ with a window that only spans the 
degenerate HOMO and LUMO pairs ($\pi$ and $\pi^*$ orbitals). In the latter case all other 
occupied and virtual orbitals are shifted with the HOMO and LUMO correction, respectively. 
Both partners of each pair are corrected explicitly, since the shift of a degenerate 
partner outside the window was found to be unreliable. 

Over most of the range the difference between the gradients with full and with fixed 
corrections, shown in the lower panel, is small. For ev$GW$ it decreases smoothly and 
monotonically. It is negative because the corrected HOMO--LUMO gap decreases with the 
bond length. For $G_0W_0$ with the standard window the 
contribution is similar up to 1.40~\AA, 
but irregular beyond,\cite{Veril2018} where it varies between $+20\times10^{-3}$ and 
$-55\times10^{-3}\,E_\mathrm{h}/a_0$, flipping sign and being two to six times larger 
than the ev$GW$ contribution. With the HOMO/LUMO window, in contrast, the contribution is 
smooth over the whole range ($+4.0\times10^{-3}$ at 1.04~\AA\ to 
$-5.9\times10^{-3}\,E_\mathrm{h}/a_0$ at 1.52~\AA). The individual corrections of the 
other orbitals are therefore the origin of the irregularity.
The irregularity is a near-singularity of the quasiparticle equations of $G_0W_0$. 
Where the renormalization factor $Z_p$ is far from unity, the gradient contribution spikes. 
This happens at approximately 1.45~\AA, where $Z_p$ of the fourth orbital drops to values 
of 0.003 and below. Close to such points the agreement of the analytic gradients with finite 
differences also worsens significantly, but both show the same qualitative behavior. 
The problem becomes more severe the more orbitals are corrected, as such a point is then 
encountered more frequently. The HOMO and LUMO were free of it in the present test, but it is 
not guaranteed that no singularity will be hit,\cite{vanSetten2013} and for core orbitals 
the issue is likely to become more severe. Contrary, ev$GW$ solves the quasiparticle equations 
iteratively and does not rely on a single-step linearization, so the issue is absent. 
This explains the increased stability of ev$GW$-BSE gradients compared to their 
$G_0W_0$-BSE counterparts.

For the practical question of the approximation, the position of the minimum is decisive. 
The gradient with fixed corrections vanishes at 1.319~\AA\ for all three variants. 
Including the contribution shifts the stationary point by $+0.003$~\AA\ 
($G_0W_0$, both windows) and $+0.006$~\AA\ (ev$GW$), so that the approximation shortens 
the C--C bond by 0.003 to 0.006~\AA. This is of the size of the average lengthening of the heavy-atom bonds 
by the full gradient in Section~\ref{sec:geometry}, and accounts for a considerable part of the systematic 
underestimation of the bond lengths with fixed corrections. At the same time, it is below the RMSD of the tested 
$GW$-BSE variants and below the RMS error of about 0.01~\AA\ reported for excited-state bond 
lengths from hybrid TD-DFT relative to CC2.\cite{Wang.Durbeej:How.2020}
The approximation is thus small in absolute terms but not negligible for a systematic bias of this size.

\subsection{Charge-transfer state of DMABN}
\label{sec:dmabn}

As a first test beyond the valence excitations of the benchmark set, 
the lowest charge-transfer (CT) state of 4-(dimethylamino)benzonitrile 
(DMABN) was optimized in its twisted (TICT) form,\cite{Grabowski2003,Rappoport2004}
in which the amino group is perpendicular to the ring.
The ground-state and CT minima optimized with EOM-CCSD 
in the cc-pVDZ basis by Knysh~et~al.\cite{Knysh2022} serve as reference.
The optimizations used a PBE0 reference, the full quasiparticle gradient 
for both $GW$ variants, and were started from the reference CT structure.
Both converged in seven cycles, giving some hope in actual geometry
optimizations being less troublesome than the small molecule test set.

\begin{figure}[htbp]
	\centering
	\includegraphics[width=\columnwidth]{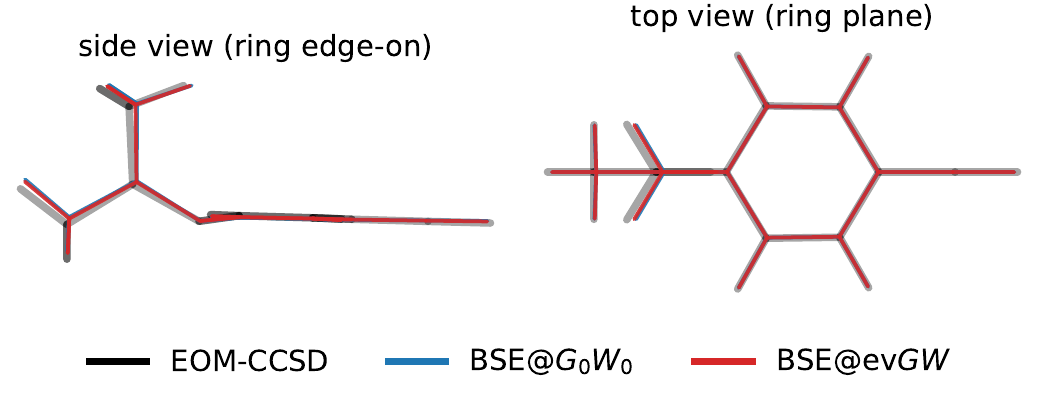}
	\caption{Overlay of the TICT minimum of DMABN from EOM-CCSD (black, wide) and from
		BSE@$G_0W_0$ (blue) and BSE@ev$GW$ (red) with the full quasiparticle gradient and a PBE0 reference.
		The structures were aligned on the ring and the N and cyano atoms.}
	\label{fig:dmabn}
\end{figure}

Figure~\ref{fig:dmabn} overlays the three structures.
Both $GW$ variants retain the perpendicular arrangement
(twist angle between the amino and ring planes of $90^\circ$, as in the reference)
and the planar amino nitrogen, and the root-mean-square deviation 
of the heavy atoms from the reference is
0.07~\AA\ ($G_0W_0$) and 0.06~\AA\ (ev$GW$) after alignment.
The characteristic bond pattern of the CT state, i.e., 
the elongation of the N--C(ipso) and C(ipso)--C(ortho) bonds
and the shortening of C(ortho)--C(meta) relative to the ground state, 
is reproduced in direction. Compared to EOM-CCSD the changes are however damped.
The N--C(ipso) bond increases from 1.391~\AA\ in the ground state to 1.431~\AA\ in the reference,
but only to 1.410 ($G_0W_0$) and 1.404~\AA\ (ev$GW$),
and the C(ipso)--C(ortho) bond from 1.421 to 1.460, 1.445 and 1.448~\AA, respectively.
The largest deviation concerns the cyano group, where the C(para)--CN and C$\equiv$N bonds are
0.03 and 0.02~\AA\ shorter than in the reference 
(1.405/1.403 vs.\ 1.436~\AA\ and 1.158 vs.\ 1.177~\AA)
and, in the case of C$\equiv$N, even shorter than in the ground state.
The two $GW$ variants differ by less than 0.006~\AA\ in every bond.
The underestimation of bond elongations is consistent with the trend found for the benchmark set,
however EOM-CCSD at a double-$\zeta$ basis is likely not converged, 
therefore the differences in this single example might overemphasize 
actual deviations. The main conclusion remains unaffected: starting from the reference structure, 
$GW$-BSE optimizations converge to a charge-transfer minimum with the correct qualitative geometry.

%%%%%%%%
\section{Conclusions}
%%%%%%%%%%%%%%%%%%%%%%%%%%%%%%%%%%%%%%%%%%%%%%%%%%%%%%%%%%%%%%%%%%%%%
Analytic BSE excited-state gradients have been formulated and implemented
within the RI approximation. The response of the static screening is mapped onto
the techniques of RI-RPA gradients and enters the TDDFT Lagrangian through
additional densities. The cost scales as $\mathcal{O}(N^4)$ with batched integral processing.
The derivative of the $GW$ corrections can be included exactly through one additional adjoint equation,
at an estimated cost of three to four times the preceding RI-$GW$ calculation.

In excited-state optimizations of the 0--0 benchmark set of Loos and Jacquemin,
BSE with a PBE0 reference and fixed corrections is of the same quality as TD-DFT for the heavy-atom bonds of singlet states,
and closer for the X--H bonds, while eigenvalue self-consistency mainly makes the optimizations more robust.
The full quasiparticle gradient removes much of the systematic underestimation of the bond lengths,
particularly for BHLYP, and reduces the dependence on the functional.
It does so at the price of the advantage for the X--H bonds with PBE0 and of
failing optimizations for some triplet states.
For the twisted intramolecular charge-transfer state of DMABN, 
optimizations with the full gradient started from the EOM-CCSD structure 
converge to a minimum that retains the perpendicular twist and the planar amino group
of the charge-transfer state for both variants. The changes of the ring bond lengths 
upon excitation have the correct direction.
A systematic assessment for charge-transfer states, the main motivation for the method, 
remains to be done.

%%%%%%%%%%%%%%%%%%%%%%%%%%%%%%%%%%%%%%%%%%%%%%%%%%%%%%%%%%%%%%%%%%%%%

%\begin{acknowledgement}

%\end{acknowledgement}

\bibliography{literature}

\end{document}